\pdfoutput=1
\documentclass[11pt]{article}

\usepackage{acl}

\usepackage{color, soul}
\usepackage{times}
\usepackage{latexsym}
\usepackage{graphicx}
\usepackage{subcaption}
\usepackage[normalem]{ulem}
\usepackage{booktabs}
\useunder{\uline}{\ul}{}

\usepackage[T1]{fontenc}
\usepackage[utf8]{inputenc}

\usepackage{microtype}

\AtBeginDocument{%
  }
    	
\title{Seasonality Based Reranking of E-commerce Autocomplete Using Natural Language Queries}

\author{Prateek Verma \\
  Walmart Global Tech \\ Bellevue, WA  \\
  \texttt{prateek.verma@walmart.com} \And
  Shan Zhong \\
  Walmart Global Tech \\ Bellevue, WA \\
  \texttt{shan.zhong@walmart.com} \\\AND
  Xiaoyu Liu \\
  Walmart Global Tech \\ Hoboken, NJ \\
  \texttt{xiaoyu.liu@walmart.com} \And
  Adithya Rajan \\
  Walmart Global Tech \\ Hoboken, NJ \\
  \texttt{adithya.rajan@walmart.com}\\} 

\begin{document}
\maketitle
\begin{abstract}
Query autocomplete (QAC) also known as typeahead, suggests list of complete queries as user types prefix in the search box. It is one of the key features of modern search engines specially in e-commerce. One of the goals of typeahead is to suggest relevant queries to users which are seasonally important. In this paper we propose a neural network based natural language processing (NLP) algorithm to incorporate seasonality as a signal and present end to end evaluation of the QAC ranking model. Incorporating seasonality into autocomplete ranking model can improve autocomplete relevance and business metric.
\end{abstract}

\section{Introduction}

Query autocomplete (QAC) is a common feature in most modern search engines. It refers to the task of suggesting complete queries given the prefix consisting of limited number of characters \cite{cai2016survey}. Goal of QAC is to help users formulate their query or predict user’s intent. Studies have shown that it can significantly reduce the number of characters typed, thus reducing physical and cognitive overload \citep{shokouhi2012time, zhang2015adaqac}. An autocomplete system typically consists of two steps: \emph{Matching} and \emph{Ranking}: Matching refers to the task of generating candidates using some form of string matching or retrieval techniques using data structures such as prefix trees. In the second step (Ranking) the matched suggestions are ordered according to their expected likelihood \cite{shokouhi2013learning}. 

The focus of this paper is \emph{Ranking} step, specifically, generation of the seasonality scores for each query such that it can be generated offline and incorporated in the autocomplete ranking model. Seasonality plays an important role in e-commerce search. Likelihood of users searching for seasonal queries in that particular season is higher than the non-seasonal queries. For example, in winter, users are more likely to search for winter related queries like \emph{winter gloves} or \emph{winter hats} than generic version of \emph{gloves} or \emph{hats}. Therefore it makes sense to rank them higher during winter season. Since seasonal queries is mostly searched during specific part of the year they do not have very high aggregate popularity score. So, the common practice of ranking queries by their aggregate popularity score can overshadow this temporal factor.

In this paper, we define the query's seasonality score based on query volume similar to the idea mentioned in~\cite{Yang2021} and propose a simple feed forward neural network architecture to predict score given the query and month. The model learns relationship between query tokens and the time period specified in the training data and can predict seasonality scores for new queries. This score can be generated for all the queries regardless of whether it is head, torso or tail and can be computed offline. It can then be integrated in the QAC ranking model to reflect seasonality in the suggested queries and their ranking. We also show the improvement in relevance and business metrics using offline evaluation and online experiments.

\let\thefootnote\relax\footnote{Accepted at The Sixth Workshop on e-Commerce and NLP (ECNLP 6), KDD'23, Long Beach, CA}

\section{Related Works}

\citet{cai2016survey} reviewed and summarized recent advancements over the auto-complete works in their survey paper. One of the most common approach for ranking QAC suggestions is to $aggregate$ the query frequency over query log and use these for ranking. \citet{bar2011context} referred to this as \emph{MostPopularCompletion} (MPC).
\begin{equation}
	MPC(p) = \arg \max_{q \in C(p)}w(q), 
	w(q) = \frac{f(q)}{\sum_{i \in Q}f(i)},
\end{equation}
where $f(q)$ denotes the number of times the query $q$ appears in the search log $Q$ and $C(p)$ is a set of query completions that start with the prefix $p$. Under the MPC model, the candidates are ranked by their past popularity and the scores do not change as long as the same query log $Q$ is used. It is assumed here that the current query popularity distribution will remain the same as that previously observed.

Time sensitive QAC approaches have been studied in \citep{shokouhi2012time, whiting2014recent, strizhevskaya2012actualization}. \citet{shokouhi2012time} analyzed why it is hard for MostPopularCompletion (MPC) based approach to handle queries whose popularity may change over a period of time. To account for such temporal behavior, they propose a time sensitive QAC ranking model. Here the default aggregate candidate scores are replaced with forecasted values computed by time-series models based on query history. In \cite{whiting2014recent}, the author uses predicted popularity based on short-range time period regression models  to rank the suggestions. In addition, \citet{jiang2017exploiting} proposed a hybrid QAC model combining periodicity with burst trend to predict future popularity of queries. They use Discrete Fourier Transform to find the periodic behavior of each query’s popularity. \citet{cai2014time} proposed a hybrid QAC model which considers predicted query popularity and user-specific context. The predicted query popularity is defined based on long-term time series analysis plus recent trend based on regression model.

Seasonality in queries has been studied before. \citet{shokouhi2011detecting} used time series decomposition techniques to identify seasonal queries in the context of web search. \citet{vlachos2004identifying} built time series for each query and phrase in the query logs and detected significant periodicities and burst in sequences. A recent work on seasonal relevance in ecommerce by \citet{Yang2021} leverages deep learning based language modeling task to learn and assign seasonal relevance scores to items. These scores are used as signals in search ranking. 

In this paper, our work focuses on time-sensitive auto-completion in QAC ranking. The primary difference between prior work and our work is that we use a neural network model to assign seasonality score to queries. Benefit of the deep learning based NLP approach is that it can learn from large amount of queries present in the search logs and learn to model the relationship between query tokens and time of the year (month in our case). This enables the feasibility to predict the seasonality for unseen or rare queries with limited historical data. E-commerce search logs typically follow long tail distribution, and such characteristics make language model based deep learning techniques more desirable since the queries in QAC will have comprehensive coverage for the seasonality score.

Application of deep learning in the context of QAC has been studied before such as \citep{park2017neural, yuan2021deep}. However, the direction of work is orthogonal to ours. \citet{park2017neural} proposed a neural language model to generate a queries for unseen or rare prefixes, and \citet{yuan2021deep} proposed deep pairwise learning to rank model that employs contextual and behavioral features to rank the queries by minimizing pairwise loss. 

\section{Seasonality}

We conduct our study for the queries in our search logs. Queries tend to show seasonality patterns, which is demonstrated by the peak and off-peak traffic during different periods of the year. 
For example, in November and December queries related to Thanksgiving and Christmas have higher traffic. In auto-complete, our goal is to assign a higher weight to these queries according to their seasonalities. 

\subsection{Definition of seasonality} \label{seasonality1}

Similar to \cite{Yang2021} we define seasonality of a query in a given month as probability of seeing the query in that month conditioned on its occurrence. This can be estimated by the query traffic or query volume V as:

\begin{equation} 
	V_{qm} = \frac{\frac{t_{qm}}{t_{m}}}{\sum_{m^{'}=1}^{12}\frac{t_{qm^{'}}}{t_{m^{'}}}}, \label{eq:1}
\end{equation}
where $t_{qm}$ denotes the traffic of query $q$ in month $m$, and $t_m$ is the overall traffic of the month.

An advantage of the above equation is that, it enforces the value to be between 0 and 1 for each query - month pair. This allows us to compare seasonality component of each query independently, regardless of the aggregate popularity of the query. Also, normalizing query volume in the given month $t_{qm}$ by total query volume in that month $t_m$ takes care to a certain extent of changing consumer shopping pattern overall or external circumstances across different time period in the year.

\begin{figure}[h]
	\centering
	\includegraphics[width=\linewidth]{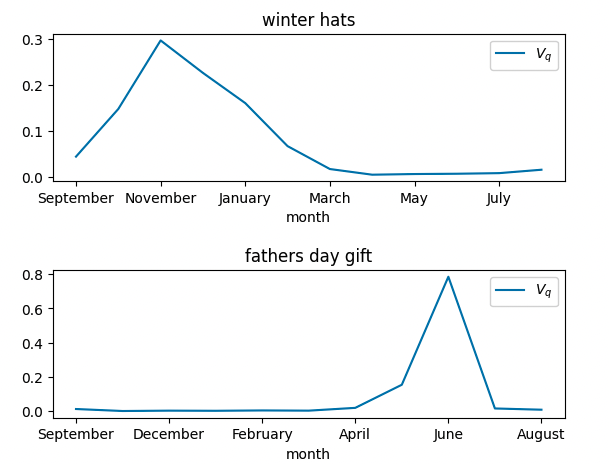}
	%\Description{Line graph for seasonality}
	\caption{Sample query volume  $V_{qm}$ across the twelve months}
	\label{fig:linegraph}
\end{figure}

Figure \ref{fig:linegraph} indicates the trend that is captured by the above Eq. (\ref{eq:1}) in the year for sample queries \emph{winter hats} and \emph{fathers day gift}.

\section{Modeling seasonality}

Seasonality for a given query $q$ in the month $m$ can be estimated from data through \begin{math}V_{qm}\end{math} . However, it has couple of limitations. First, it can only be computed for queries which have traffic information available. Second, it can be noisy due to factors like the query being tail/torso, sparse data, and other external scenarios.

To take care of cold start and reduce noise due to the above mentioned issues, we build a neural network to model the seasonality values based on the query text $q$, month $m$ and \begin{math}V_{qm}\end{math} which is defined in the previous section. $q$ and $m$ is the input to the neural network and \begin{math}V_{qm}\end{math} is the target value. The model can be represented by a function  $f(\theta): $  ($q$, $m$) $\rightarrow$ \begin{math}S_{qm}\end{math}, and
$f(\theta)$ is learnt by minimizing the mean square error 
\begin{equation}
	L = \sum_{q\in D}(S_{qm} - V_{qm})^2
	\label{equation2}
\end{equation}
Here $D$ is the dataset of training documents. $S_{qm}$ is the value predicted by the model, month $m$ is a value between 1 and 12 representing January to December. It is represented as a one hot encoding. Query $q$ is the embedding of query text. We model the function $f$ using a feed forward neural network. 

To mitigate the issue of disambiguating new product launches with seasonality, mutiple years can be considered for the training data.

\subsection{Training Dataset Generation}

We use search query logs to create the training dataset. It is a log of every single query the search platform has seen in the past. Firstly, we compute monthly query volume for each query $t_{qm}$ and filter out queries that have query volume below certain threshold say $K$. Then we aggregate monthly query volume of all search queries $t_{m}$. Finally, using the Eq. (\ref{eq:1}) we compute \begin{math}V_{qm}\end{math}. This is normalized and scaled to a value between 0 to 1 using monthly and yearly search traffic. After the thresholding and some random sampling, our training dataset consist of about 337,000 distinct queries.

\begin{figure}[!h]
	\centering
	\includegraphics[width=\linewidth,scale=0.5]{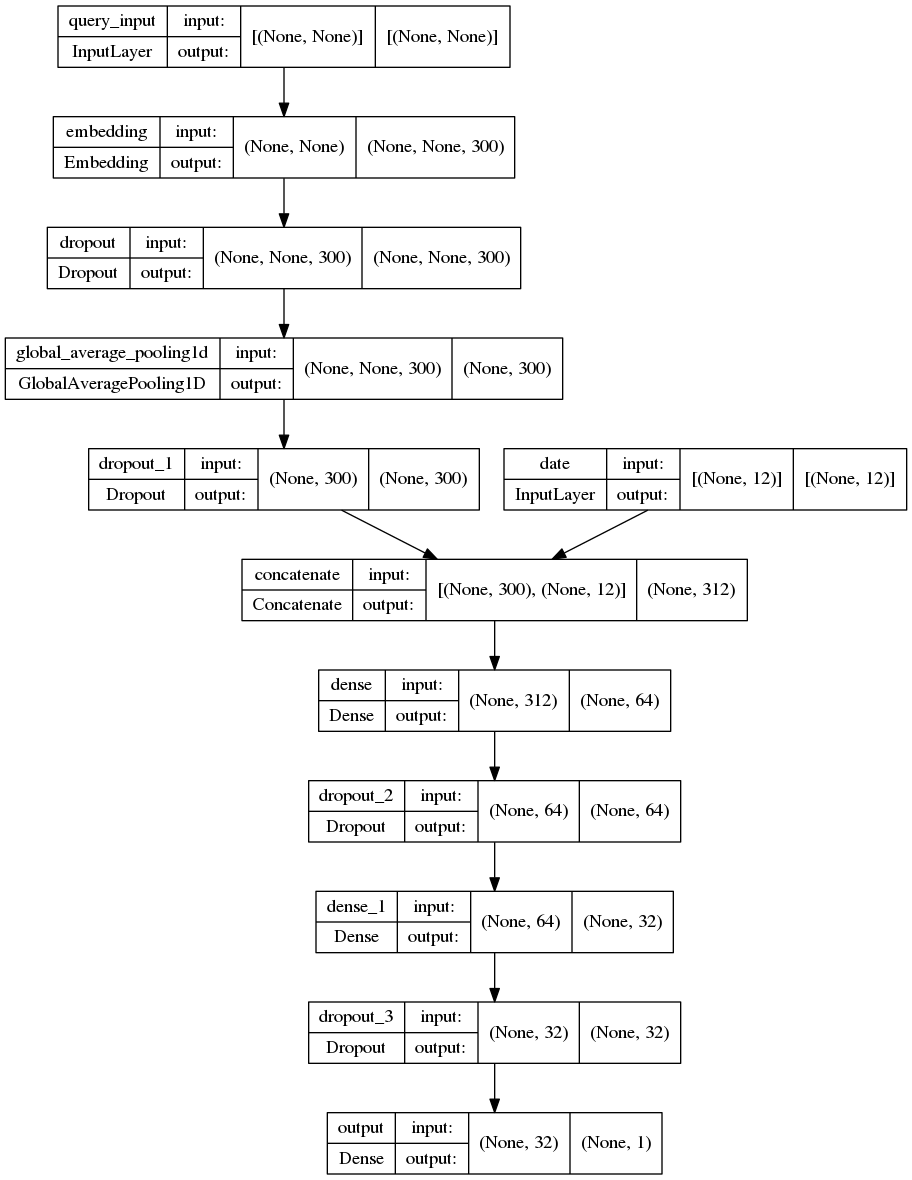}
	\caption{Feed-forward neural network for predicting seasonality value for a given query and month}
	\label{fig:networkarchitecture}
	
\end{figure}

\subsection{Model Architecture}
Figure~\ref{fig:networkarchitecture} shows the model architecture. It accepts query $q$ and month $m$ as inputs. We encode query using dense vector representation \cite{mikolov2013distributed} and month by a one-hot vector representation. We initialize the embedding layer with a 300 dimensional pre-trained Glove embedding \cite{pennington2014glove}. This layer is set as trainable. 

We use dropout ratio of 0.2 in between the layers to reduce overfitting, it was obtained via hyper parameter tuning. The layers in between use $relu$ as non-linear activation function. Finally, the output layer is a single neuron with $linear$ activation function. 
The intuition behind the architecture is to capture relevancy of the given query and month for building a regression model which can predict \begin{math}V_{qm}\end{math}. To select number of layers, dropout and the dimensions of Glove embedding, we performed hyper parameter tuning and used \emph{mean square error} as the evaluation metric. We were able to obtain optimal \emph{mean square error} using the configuration of the architecture as described above. 

$Adam$ $optimizer$ and $mean$ $square$ $error$ loss function are used for training the model.

In table~\ref{table:lossmetrics}, we show the loss metric for 100 and 300 dimensional Glove embedding pretrained with 6B and 42B tokens \cite{pennington2014glove}. We report the numbers for configuration of the dropout parameters for which we obtained the best \emph{mse} loss.

The neural network model can be further improved by potentially introducing attention layer. Another way to model seasonality is to build a sequence to sequence deep learning model and have it predict twelve numbers for a given query which correspond to seasonality values for the twelve months in a year. We leave the evaluation of these methods as future work.
Note that the objective of this paper is to define seasonality and evaluate if integrating it with the auto-complete ranking model improves the overall ranking performance. 

\begin{table}[]
	\centering
	\caption{Test error to evaluate Glove embeddings}
	\label{table:lossmetrics}
	\begin{tabular}{cc}
		\hline
		\textbf{Glove Embedding} & \textbf{Test Error (MSE)} \\
		\hline
		6B 100D         & 0.00218          \\
		6B 300D         & 0.00241          \\
		42B 300D        & 0.00215         \\
		\hline
	\end{tabular}
\end{table}

\begin{table*}[]
	\centering
	\caption{Autocomplete suggestion for the prefix "memo" a week before memorial day}
	\label{table:example_ranking}
	\begin{tabular}{cc}
		\hline
		\textbf{Control} & \textbf{Test Group}  \\
		\hline
		memory foam mattress topper & \textcolor{red} {memorial day flowers}    \\
		memory foam mattress        & memory foam mattress topper \\
		memory foam pillow          & memory foam mattress        \\
		\textcolor{red} {memorial day flowers}        & \textcolor{red} {memorial day decorations}    \\
		memory card                 & memory foam pillow          \\
		\textcolor{red} {memorial day decorations}    & \textcolor{red} {memorial day}                \\
		memory foam futon           &\textcolor{red} {memorial flowers}       \\    
		\hline               
	\end{tabular}
\end{table*}

\begin{table*}
	\centering
	\caption{Offline evaluation: mean reciprocal rank on four different platforms}
	\label{table:offline_evaluation}
	\begin{tabular}{ccccc}
		\hline
		\textbf{Device} & \textbf{Desktop} & \textbf{Mobile web} & \textbf{IOS} & \textbf{Android}  \\
		\hline
		MRR lift & +0.37\%  & +0.96\% & +0.24\% & +0.13\%  \\
		\hline
	\end{tabular}
\end{table*}

\section{Experiment}

In this section, we discuss the impact of integrating seasonality into the auto-complete ranking model. Our analysis is based on seasonality values modeled using the predictive approach described in the previous section. We study the impact through offline evaluation and online experiment.

\subsection{System Implementation for QAC ranking model}
Our auto complete system consists of two components: \emph{L1} and \emph{L2} ranker (a re-ranking model). \emph{L1} ranker is linear time series model trained on a historical search log. It is trained to predict a linear combination of number of add-to-carts, clicks and impressions a given query would receive in upcoming weeks. This runs offline in a batch manner and helps us select initial corpus of queries which we index into our auto-complete platform. \emph{L2} ranker is a runtime re-rank model: a function of the offline score generated by \emph{L1} ranker and the user session level feature. It performs re-ranking and shows top ten suggestion to users. This function could be a simple linear model or a more sophisticated  function estimated through deep learning models. In \emph{L2} ranker we incorporate the seasonality value predicted by the deep learning model as a signal.

The \emph{L1} ranker already captures seasonality and time sensitivity to a certain extent. By integrating the seasonality signal into the re-rank model \emph{L2}, we are interested to observe the additional improvements to the ranking. 
We obtained the weight for the seasonality signal in the L2 ranker by fine-tuning it on evaluation datasets. For offline and online evaluation, the control group is the QAC model where \emph{L2} ranker does not have seasonality signal and the test group is the QAC model where \emph{L2} ranker includes the seasonality signal.

 Table~\ref{table:example_ranking} shows ranking of QAC suggestions for the prefix "memo" a week before the memorial day in the US. We can see suggestions related to the memorial day event is ranked higher in test compared to the control goup. It is to be noted that extra suggestions relevant to the event show up because we re-rank top \emph{N} suggestions obtained from L1 ranker to display top \emph{K} to the user, where \emph{N} > \emph{K}.

\subsection{Offline evaluation}
The offline evaluation system uses queries from historical search logs to query the auto-complete system and records the suggestions for each character of a given query. From the search logs we know the ground truth query i.e. final query user had submitted. Thus we can compute reciprocal rank for each search request. We use this technique to compute \emph{mean reciprocal rank} (MRR) from about 400,000 prefixes generated from randomly sampled 50,000 search queries for four platforms: desktop, mobile web, IOS and Android. Table~\ref{table:offline_evaluation} shows the impact of incorporating seasonality feature into the autocomplete ranking model on MRR for the four platforms. We observe statistically significant lift in MRR.

\subsection{Online experimentation}
We conduct an online A/B test and study the effect of incorporating seasonality feature into our autocomplete ranking. Experiment shows the ranking model with seasonality signal in iOS and android yields a lift of 0.25\% (p-value < 0.1) and 0.34\% (p-value < 0.1) in GMV (Gross Merchandise Value) respectively.

\section{Conclusion and Future Work}
In this work, we introduced a method to make auto-complete suggestions seasonally relevant using deep learning based NLP algorithm to predict seasonality for a given query and month. The predicted score is integrated as a feature in the auto-complete learning-to-rank model. We also present end-to-end evaluation of auto-complete ranking model and measure the gains in \emph{mean reciprocal rank} (MRR).

As future work, we intend to explore following areas: First, generation of seasonality score at a more granular level such as biweekly, weekly intervals or include day of the week as signal in the ranking model. Second, making use of transformer based architecture to model the seasonality of queries. Third, study the interaction of seasonality and query category, and include seasonality at the category level in the learning-to-rank model. 

\section*{Acknowledgements}
We would like to thank Ari Kast, Aishwarya, Sanjay Shah and Kevin Li from the Walmart Search Team for helping with the platform engineering support. We are also very thankful to John Yan for providing leadership support.

\bibliography{seasonality_references}

\end{document}